\documentclass[preprintnumbers,amsmath,amssymb,floatfix]{revtex4}
\usepackage{amsmath,graphicx,amsfonts}
\usepackage{dcolumn}

\newcommand{\be}{\begin{equation}}
\newcommand{\ee}{\end{equation}}

\begin{document}

\title{Adatom interaction effects in surface diffusion}
\author{Yuri B. Gaididei}
\affiliation{Bogolyubov Institute for Theoretical Physics,
Metrologichna str. 14 B, 03680, Kiev, Ukraine}
\author{Vadim M. Loktev}
\affiliation{Bogolyubov Institute for Theoretical Physics,
Metrologichna str. 14 B, 03680, Kiev, Ukraine}
\author{Anton G. Naumovets}
\affiliation{Institute of Physics,
Prospect Nauki 46, 03680, Kiev, Ukraine}
\author{Anatoly G. Zagorodny}
\affiliation{Bogolyubov Institute for Theoretical Physics,
Metrologichna st. 14 B, 03680, Kiev, Ukraine}

\date{\today}

%%%%%%%%%%%%%%%%%%%%%%%%%%%%%%%%%%%%%%%%%%%%%%%%%%%%%%%%%%%%%%%%%%%%%70
%
%         ABSTRACT
%
%%%%%%%%%%%%%%%%%%%%%%%%%%%%%%%%%%%%%%%%%%%%%%%%%%%%%%%%%%%%%%%%%%%%%70

\begin{abstract}
Motivated by recent research  of Nikitin {\it et al} (J.Phys.D {\bf 49},055301(2009)), we examine the effects of interatomic  interactions on adatom surface diffusion. By using a mean-field approach in the random walk problem, we derive a nonlinear diffusion equation and analyze its solutions. The results of our analysis are in good agreement with direct numerical simulations of the corresponding discrete  model. It is shown that by analyzing a time dependence of  adatom concentration profiles one can estimate the type and strength of interatomic interactions.

\end{abstract}

%\pacs{75.10.Hk, 75.40.Mg, 05.45.-a, 72.25.Ba, 85.75.-d}

% 75.10.Hk    Classical spin models
% 75.40.Mg    Numerical simulation studies
% 05.45.-a    Nonlinear dynamics and nonlinear dynamical systems.
% 72.25.Ba    Spin polarized transport in metals
% 85.75.-d    Magnetoelectronics; spintronics: devices exploiting spin polarized transport or integrated magnetic fields

%%%%%%%%%%%%%%%%%%%%%%%%%%%%%%%%%%%%%%%%%%%%%%%%%%%%%%%%%%%%%%%%%%%%%70

\maketitle
\section{Introduction}
Diffusion is ubiquitous in Nature.  It determines the behavior and controls the efficiency  of many
 biological and technological processes.  Examples include the wetting, conductivity of biological membranes, catalysis, growth of crystals, sintering, soldering {\it etc}.  Surface diffusion  is particularly important in nano-technological processes which are aimed to obtain objects of submicron  sizes where the surface properties are of the same importance as  the bulk ones. Macroscopic description of diffusion is based on Fick's law, which postulates proportionality between the particle flux and the concentration gradient.   Establishing a link between macroscopic laws of diffusion and microscopic non-equilibrium density matrix approach is one of the most challenging and important problems of non-equilibrium statistical mechanics (for reviews see, e.g. \cite{gomer,naumovets_rev,gouyet,alanissila,naumovets2005,antezak}). Surface diffusion is essentially many-particle process.
 Even at very low coverage when the interaction between adatoms is negligible, the random walk of an isolated  adsorbed particle is a collective motion due  its  interaction with substrate atoms \cite{alanissila,naumovets2005}. The walker moves in a potential landscape which is changed by the walker  \cite{lam}. The walk on a deformable  medium  when the walker leaves  behind a trail and due to slow relaxation the trail affects the next walker, is also a collective process \cite{gaididei,huang}.  At finite coverage particles at surfaces, in addition to the interaction with a substrate, experience lateral interactions of different origin: the attractive van der Waals, oscillating electronic exchange and multipole-multipole electrostatic interactions. The dipole-dipole interaction which is due to a polar (mainly dipolar) character of adsorption bonds is long-ranged (as $r^{-3}$) and generally repulsive since all dipole moments are essentially parallel (see  review papers, e.g. \cite{alanissila,naumovets2005}).

 Bowker and King \cite{bowker} used  Monte-Carlo simulations in order to clarify the effect of lateral interactions of adatoms on the shape of evolving concentration profiles in surface diffusion. They showed that the intersection point of the diffusion profiles with the initial  stepwise profile lies above $\theta_{max}/2$ in the case of lateral repulsion and below $\theta_{max}/2$ in the case of attraction ($\theta_{max}$ is the maximum concentration in the initial step).

 In a quite recent paper \cite{nikitin} an approach based on error function expansion  was proposed to fit experimental concentration profiles. This algorithm provides a high-accuracy fitting and allows  extracting the concentration dependence of diffusivity from experimental data.   The goal of our paper is to model and examine  the effects of interatomic  interactions on adatom surface diffusion. Starting with nonlinear random walk equations where the interatomic interactions are considered in the mean-field approach, we derive a nonlinear diffusion equation and analyze its solutions. The results of our analysis are in good agreement with direct numerical simulations of the corresponding discrete  model. It is shown that by analyzing a time dependence of  adatom concentration profiles one can estimate the type and strength of interatomic interactions.
 The paper is organized as follows. In Sec.1 we present the model. In Sec.2  we study both analytically and numerically interaction effects for the case of low adatom coverage. Sec.3 is devoted to analytical treatment of nonlinear diffusion in the case of non-monotonic concentration dependence of the diffusion coefficient. We also compare our results with the results of full scale numerical simulations and results of experimental observation.
 Sec.4 presents some concluding remarks.

\section{Model and equations of motion}

The transport of particles on the surface is described by the set of random walk equations
\begin{eqnarray}\label{random_walk_eqn}
\frac{d}{dt}\,\theta_{\vec{n}}(t)=\sum_{\vec{\rho}}\,\Big[W_{\vec{n}+\vec{\rho}\rightarrow\vec{n}}\,
\theta_{\vec{n}+\vec{\rho}}(t)\,
\Big(1-\theta_{\vec{n}}(t)\Big)-
W_{\vec{n}\rightarrow\vec{n}+\vec{\rho}}\,\theta_{\vec{n}}(t)\,\Big(1-\theta_{\vec{n}+\vec{\rho}}(t)\Big)\Big]
\end{eqnarray}
where $\theta_{\vec{n}}$ is the probability for a particle occupying the $\vec{n}$-th binding site on the surface (in the literature on surface science this quantity has a meaning of coverage), $W_{\vec{n}\rightarrow\vec{n}+\vec{\rho}}$ gives the rate of the jumps from the binding site $\vec{n}$ to a neighboring  site $\vec{n}+\vec{\rho}$ (the vector $\vec{\rho}$ connects nearest neighbors) . The terms $\Big(1-\theta_{\vec{n}}(t)\Big)$ in Eqs. (\ref{random_walk_eqn}) take into account the fact that there may be only one adatom at a given site or, in other words, a so-called kinematic interaction. The  probability with which the particle jumps from site $\vec{n}$ to a nearest neighbor $\vec{n}+\vec{\rho}$  satisfies  the detailed balance condition,
\begin{eqnarray}\label{det-rate}
W_{\vec{n}\rightarrow\vec{n}+\vec{\rho}}\,e^{-\beta E_{\vec{n}}}=W_{\vec{n}+\vec{\rho}\rightarrow\vec{n}}\,e^{-\beta E_{\vec{n}+\vec{\rho}}},
\end{eqnarray}
where  $E_{\vec{n}}$ is the binding energy of the particle situated at  site $\vec{n}$ ,  $\beta=1/k_B T, ~k_B$ is the Boltzmann constant and $T$ is the temperature of the system. For the transition rates we choose
\begin{eqnarray}\label{rate}W_{\vec{n}\rightarrow\vec{n}+\vec{\rho}}=w_{\vec{\rho}}\,e^{\beta E_{\vec{n}}}\end{eqnarray}
 which corresponds to setting the activation energy for a jump  to the initial binding energy. Here $ w_{\vec{\rho}}=\nu_{\vec{\rho}}\,e^{-\beta E_b} \,\,(w_{\vec{\rho}}=w_{-\vec{\rho}})$ is  the jump rate of an isolated particle with standard notations: $\nu_{\vec{\rho}}$  is a frequency factor and $E_b-E_{\vec{n}}$ is the height of the random walk barrier. Inserting Eq. (\ref{rate})  into Eqs. (\ref{random_walk_eqn}) we obtain that the random walk of particles on the surface is described by a set of equations
 \begin{eqnarray}\label{random_walk_eqn_mod}\frac{d}{dt}\,\theta_{\vec{n}}(t)=
 \sum_{\vec{\rho}}\,w_{\vec{\rho}}\,\Big[\Big(1-\theta_{\vec{n}}(t)\Big)
\,\theta_{\vec{n}+\vec{\rho}}(t)\,e^{\beta\,E_{\vec{n}+\vec{\rho}}}
-\,\theta_{\vec{n}}(t)\,\Big(1-\theta_{\vec{n}+\vec{\rho}}(t)\Big)\,e^{\beta\,E_{\vec{n}}}
\Big]\;.\end{eqnarray}

In the case when the characteristic size of the particle distribution inhomogeneity is much larger than the lattice spacing one can  replace $\theta_{\vec{n}}$ and $E_{\vec{n}}$ by the functions $\theta(\vec{r})$ and $E(\vec{r})$ of the continuous variable $\vec{r}$ and, by expanding the functions $\theta(\vec{r}+\vec{\rho})$ and $E(\vec{r}+\vec{\rho})$ into a Taylor series,  obtain from Eqs. (\ref{random_walk_eqn_mod})   that in the continuum approximation the transport of particles on the surface is described  by the equation of the form
\begin{eqnarray}\label{nonl_diff_eq-1}\partial_t \theta=w\,\nabla\Big\{\Big(\nabla\,\theta+\beta\,\theta\,(1-\theta)\,\nabla E\Big)\,e^{\beta E}\Big\}\end{eqnarray}
where  the notation $w=\frac{1}{2}\,\sum\limits_{\vec{\rho}}\,\vec{\rho}^2\,w_{\vec{\rho}}$ is used.

We will study the particle kinetics in the mean field approach when the binding energy $E$ is assumed to be a functional of the particle density $\theta(\vec{r},t)$: $E(\vec{r})= {\cal E}(\theta)$.
In this case Eq. (\ref{nonl_diff_eq-1}) takes a form of nonlinear diffusion equation
\begin{eqnarray}\label{nonl_diff_eq}\partial_t \theta=\,\nabla\Big\{D(\theta)\,\nabla \theta\Big\}\;,\end{eqnarray}
where
\begin{eqnarray}\label{nonl_diff_coeff}D(\theta)=w\,\Big(1+\beta\,\theta\,(1-\theta)\,\frac{\delta {\cal E}}{\delta \theta}\Big)\,e^{\beta {\cal E}}\end{eqnarray}
is a nonlinear (collective) diffusion coefficient.

\section{Surface diffusion at low coverage}
In what follows we restrict ourselves to studying particle distributions spatially homogeneous along the $y$ coordinate: $\theta(\vec{r},t)\equiv \theta(x,t)$.
We assume that initially the particles are step-like distributed
\begin{eqnarray}\label{initial}\theta(x,0)=\theta_{max}\, H(-x)\end{eqnarray}
where $H(x)$ is the Heaviside step function. By introducing a centered particle density $\xi(x,t)=\,\Big(\theta(x,t)-0.5\,\theta_{max}\Big)/\theta_{max}$, we see that the initial distribution $\xi(x,0)$ is an odd function of the spatial variable $x$. It is obvious that in the no-interaction case (${\cal E}=const$), when the diffusion equation (\ref{nonl_diff_eq})  is linear,  the antisymmetric character of the function $\xi(x,t)$ is  preserved for all $t>0$. This means that  in the case of noninteracting particles the concentration profile for each time moment  $t$  passes through the point $\Big(0,\frac{\theta_{max}}{2}\Big).$ However, interacting diffusing particles exhibit quite a different behavior. In 1969, Vedula and one of the authors detected  for the first time that concentration  profiles formed in  the process of surface diffusion of thorium on tungsten intersected the initial step-like profile at a point lying well above $\theta_{max}$ (see \cite{vedula,naumovets_rev}). Since then, similar behavior has been found for many electropositive adsorbates whose adatoms are known to interact repulsively. A recent example obtained in the case of surface diffusion of Li on the Dy-Mo~(112) surface was discussed in \cite{nikitin}.
 It is worth noticing that the above mentioned behavior was observed  even for rather low coverage: $\theta_{max}\,< 0.3$ (see Fig. 2 in \cite{nikitin}). Therefore to explain such a behavior one may assume that the binding energy $E(\vec{r})$ is linearly dependent on the particle concentration:
\begin{eqnarray}\label{mean_field}
E(\vec{r})=E_0+\int d\vec{r}'\,V(\vec{r}-\vec{r}')\,\theta(\vec{r}')\approx E_0+\theta(\vec{r})\,\int d\vec{r}'\,V(\vec{r}')
\end{eqnarray}
where $E_0$ is a site energy and $V(\vec{r}-\vec{r}')$ is an interaction parameter which includes all types of lateral interactions.  In this case the nonlinear diffusion coefficient (\ref{nonl_diff_coeff}) takes a form
\begin{eqnarray}\label{nonl_diff_coeff_mod}
D(\theta)=D^*\,\Big(1+\alpha \,\theta\,(1-\theta)\Big)\,e^{\alpha\theta}
\end{eqnarray}
where $$D^*=w\,e^{\beta\,E_0}\equiv\nu\,e^{-\beta\,(E_b-E_0)}$$ is the  diffusion coefficient for an isolated particle (or  a so-called tracer diffusion coefficient) and the dimensionless parameter $$\alpha=\beta\,V_0,~~~V_0=\int d\vec{r}'\,V(\vec{r}')$$ characterizes the strength of the lateral interaction.

\begin{figure}
\includegraphics[width=0.5\textwidth]{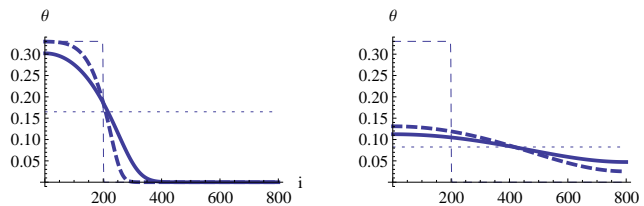}
\caption{Concentration profile obtained from nonlinear random walk equations (\ref{random_walk_eqn_1D})  for $\alpha=3$ and initial distribution given by a thin dashed curve. The left panel  shows the early stage of evolution:  $w t=500$ (solid curve), $w t=2000$ (dashed curve), the right panel shows the late stage of evolution: $~w t=40000$ (dashed),$~w t=60000$ (solid). The line which corresponds to $\theta=\theta_{max}/2$ is shown as a horizontal dotted line in the left panel and the line which corresponds to  the final state of the evolution $\theta=\theta_{max}/4$ is shown as a horizontal dotted  line in the right panel.}
\label{fig:coverage_low}
\end{figure}

The aim of this section is to develop an approach which allows  estimating the effects of interparticle interactions in the surface diffusion.
It is seen from Eqs. (\ref{nonl_diff_eq}) that the spatio-temporal behavior of the centered particle density $\xi(x,t)$  is governed by the equation
\begin{eqnarray}\label{nonl_diff_eq_2}\partial_{\tau} \xi=\,\partial_x^2\Big(\xi + P(\xi)\Big)\;, \end{eqnarray}
where $\tau=D_0\,t$  is a rescaled time and the quantity \begin{eqnarray}\label{P}P(\xi)=\frac{1}{D^{*}}\,\int\limits_0^\xi\,d\xi'\,D(\xi')-\xi\end{eqnarray} describes the nonlinear properties of the diffusion and vanishes when $\alpha\rightarrow 0$.
Taking into account that Eq. (\ref{nonl_diff_eq_2}) with the initial condition given by Eq. (\ref{initial}) is invariant under gauge transformations $\tau\rightarrow \lambda^2 \,\tau,x\rightarrow \lambda \,x,\theta\rightarrow \theta$ ($\lambda$ is an arbitrary number) one can look for a solution of Eq. (\ref{nonl_diff_eq_2})
in terms of the Boltzmann variable $z=\frac{x}{2\sqrt{\tau}}$ :  $\xi(x,\tau)=\zeta(z)$ where the function $\zeta(z)$ satisfies the equation
\begin{eqnarray}\label{nonl_diff_eq_3}\frac{d^2}{d z^2}\Big(\zeta+P(\zeta)\Big)+2 z \frac{d \zeta}{d z}=0,\end{eqnarray}
with the boundary conditions
\begin{eqnarray}\label{nonl_diff_eq_4}
\zeta(z)\rightarrow\mp\frac{1}{2},~~~z\rightarrow\pm{\infty}\;.\end{eqnarray}
Eqs. (\ref{nonl_diff_eq_3}), (\ref{nonl_diff_eq_4})  can be rewritten in the form of the following integral equation
\begin{eqnarray}\label{nonl_int_eq}\zeta(z)= \frac{\sqrt{\pi}}{4}\,\int\limits_{0}^{\infty}dz\,w_+(z)-\frac{1}{2}\Big(1- \frac{\sqrt{\pi}}{2}\,\int\limits_{0}^{\infty}dz\,w_-(z)\Big)\,\mathrm{erf}(z)
-\nonumber\\
\frac{\sqrt{\pi}}{2}\,
\int\limits_{0}^{z}dz_1\,e^{z_1^2}\,\Big(\mathrm{erf}(z)-\mathrm{erf}(z_1)\Big)\,\frac{d^2}{d z_1^2}\,P\Big(\zeta(z_1)\Big),
\nonumber\\
w_{\pm}(z)=e^{z^2}\,\Big(1-\mathrm{erf}(z)\Big)\,\frac{d^2}{d z^2}\,\Big[P\Big(\zeta(z)\Big)\pm P\Big(\zeta(-z)\Big)\Big]\;.\end{eqnarray}
where $\mathrm{erf}(z)$ is the error function \cite{abr}.
It is seen from Eq. (\ref{nonl_int_eq}) that the concentration profiles $\xi(x,t)$  for different time moments intersect at the point $\Big(0,\zeta(0)\Big)$ with
\begin{eqnarray}\label{intersect}\zeta(0)=
\frac{\sqrt{\pi}}{4}\,\int\limits_{0}^{\infty}dz\,w_+(z)\;.\end{eqnarray}
In the weak interaction/low coverage limit when $\alpha\,\theta_{max}<\,1$ one can replace the function $\zeta(z)$ in the right-hand-side of Eqs. \eqref{nonl_int_eq}, (\ref{intersect}) by its expression obtained in the linear case:$~\zeta_0(z)=\frac{1}{2}\,\mathrm{erf}(z)$ and obtain approximately that under the step-like initial condition (\ref{initial}) the concentration profiles $\theta(x,\tau)$ for different time moments intersect at the point which corresponds to the concentration
\begin{eqnarray}\label{intersect_point}\theta_0=
\Big(\frac{1}{2}+\zeta(0)\Big)\,\theta_{max}\;,\nonumber\\
\zeta(0)=\frac{\pi-2}{4 \,\pi}\,\alpha\,\theta_{max}\approx 0.1\,\alpha\,\theta_{max}\;.\end{eqnarray}
Thus the concentration  value $\theta_0$ at which the concentration profiles  intersect changes in the  presence of lateral interatomic interactions: $\theta_0>\theta_{max}/2~~(\theta_0<\theta_{max}/2)$ when the interaction is  repulsive (attractive).
We applied Eq.(\ref{intersect_point}) to analyze the results obtained in \cite{nikitin} for the diffusion of Li on the Dy-Mo (112) surface at low coverage ($\theta_{max}\approx 0.33$) for which $\theta_0\approx 0.19$ and found out that $\alpha\,\theta_{max}\approx 0.97$. It is seen that strictly speaking it is not fully legitimate to use our simple analytical  perturbation approach (which is valid for $\alpha\,\theta_{max} \ll 1$) to analyze the results of experiments \cite{nikitin} but a qualitative agreement takes place. To validate our analytical results we carried out numerical simulations of Eqs. (\ref{random_walk_eqn_mod}), (\ref{rate}) which in the 1D-case  for a system with $N$ binding sites have the form
\begin{eqnarray}\label{random_walk_eqn_1D}
\frac{d}{d\tau}\,\theta_1=(1-\theta_{1})\,e^{\beta\,{\cal E}_{2}}\,\theta_{2}-\Big(1-\theta_{2}\Big)\,
e^{\beta\,{\cal E}_{1}}\,\theta_{1},\nonumber\\
\frac{d}{d\tau}\,\theta_n=(1-\theta_{n})\,\Big(e^{\beta\,{\cal E}_{n+1}}\,\theta_{n+1}+
e^{\beta\,{\cal E}_{n-1}}\,\theta_{n-1}\Big)-\Big(2-\theta_{n+1}-\theta_{n-1}\Big)\,
e^{\beta\,{\cal E}_{n}}\,\theta_{n},
~~(n=2,... N-1),\nonumber\\
\frac{d}{d\tau}\,\theta_N=(1-\theta_{N})\,e^{\beta\,{\cal E}_{N-1}}\,\theta_{N-1}-\Big(1-\theta_{N-1}\Big)\,
e^{\beta\,{\cal E}_{N}}\,\theta_{N}
\end{eqnarray}
where $\beta \,{\cal E}_{n}=\alpha\,\theta_n$.  Thus, in our model the total number of particles is a conserved quantity. As an initial state  we used a step-like distribution \begin{eqnarray}\theta_n=\theta_{max},~~~~\mathrm{for}~~1\leq n\leq \frac{N}{4}\;,\nonumber\\
\theta_n=0,~~~~\mathrm{otherwise\;.}\end{eqnarray}We found out that for $\theta_{max}=0.33$  the concentration profiles intersect at the point $(0,0.19)$ ( as it was observed in  the experiment \cite{nikitin}) for $\alpha\approx (3\div 3.5)$ ( see Fig. \ref{fig:coverage_low})  or $\alpha\,\theta_{max}\approx (1\div 1.5)$ which is a good agreement with our analytics. Thus basing on our approach one can conclude that Li adatoms on the Dy-Mo (112) surface mostly repel each  other and the intensity of the repulsion is $V_0\approx (3\div 3.5) \,k_B\,T$.

Diffusion of Li adatoms on Dy/Mo(112) was investigated experimentally at $T=600 K$ \cite{nikitin}, so the estimated repulsion energy $V_0$ amounts to $0.16-0.18 $ eV. Let us assess this value in terms of the dipole-dipole interaction. The energy of the repulsive interaction between two dipoles having moments $p$ and situated on the surface at a distance $r$ is
\begin{eqnarray}\label{dipo-int}U_{dd}=2\,\frac{p^2}{r^3}\approx\frac{1.25\, p^2 \,[Debyes]}{r^3\,[Angstroms]}\,[eV]\end{eqnarray}
The dipole moment can be determined from the work function change $\Delta\varphi$ using the Helmholtz formula for the double electric layer:
\begin{eqnarray}\label{dipo}\mid\Delta\varphi\mid= 4\,\pi\,n\,p\,e\;,	\end{eqnarray}							
where $n$ is the surface concentration of adatoms and e is the electronic charge. For Li on the Mo(112) surface, the $p$ value at low coverage was found to be $1.4$ Debyes \cite{braun}. Then, using Eq. (\ref{dipo-int}) we can find that for two such dipoles the interaction energy $U_{dd}=0.16 $eV can be attained at a distance $r\approx 2.5 \r{A}$, which is close to the distance between the nearest adsorption sites $(2.73\r{A})$ within the atomic troughs on Mo(112). This estimation shows that the intensity of the lateral interaction deduced from the diffusion data in the way presented above seems physically reasonable. Recall, however, that $V_0$ determines a resultant effect experienced by a jumping particle from all its counterparts which, in the case of heterodiffusion, are non-uniformly distributed over the surface and provide an additional driving force (supplementary to the coverage gradient) that favors a faster diffusion of repulsing particles.

The adatom interaction effects fade away for the late stage of the evolution when the particle density becomes small  and the diffusion process transfers into a linear regime ( see Fig.  \ref{fig:coverage_low} for $t=60000$).

\section{Mean-square deviation}
The adatom interactions manifest themselves also in the integral characteristics of kinetics of adatom diffusion. It is well known that in the linear regime the  variance
\begin{eqnarray}\label{mean_sq_def}
\langle x^2\rangle=
\frac{\int\limits_{-\infty}^{\infty}\,dx\,x^2\,\theta(x,\tau)}
{\int\limits_{-\infty}^{\infty}\,dx\,\,\theta(x,\tau)}\end{eqnarray}
 behaves (in one-dimensional case) as $\langle x^2\rangle=2\tau$. Therefore  it is naturally to introduce
 a variance rate
\begin{eqnarray}\label{mean_sq_rate}
\Delta(\tau)=\Big(2-\frac{d}{d \tau}\langle x^2\rangle\Big)^2
\end{eqnarray}
 whose time dependence   provides a useful information about nonlinear effects in the diffusion process.

For this quantity   we obtain from Eq. (\ref{nonl_diff_eq_4}) that
\begin{eqnarray}\label{mean_sq_eq}\Delta(\tau)=\alpha^2\,\Big(\int\limits_{-\infty}^{\infty}\,dx\,
\,\theta^2(x,\tau)\Big{/}\int\limits_{-\infty}^{\infty}\,dx\,\,\theta(x,\tau)\Big)^2\end{eqnarray}
Assuming that initially the particles are concentrated in a finite domain in a $\Pi$-like form:
 \begin{eqnarray}\label{initial1}\theta(x,0)=\theta_{max}\,\Big(H(x+l)-H(x-l)\Big)\end{eqnarray}
 where $2 l$ is the size of the initial domain, for small nonlinearities $\alpha$ and low coverage $\theta_{max}\ll 1$  we obtain approximately that
\begin{eqnarray}\label{mean_sq_eq_ap}\Delta(\tau)=\alpha^2\,\Big(\int\limits_{-\infty}^{\infty}\,dx\,
\,\theta_{lin}^2(x,\tau)\Big{/}\int\limits_{-\infty}^{\infty}\,dx\,\,\theta_{lin}(x,\tau)\Big)^2\end{eqnarray}
where
\begin{eqnarray}\label{theta0}\theta_{lin}(x,\tau)=\frac{\theta_{max}}{2}\,\Big[\mathrm{erf} \Big(\frac{l-x}{2\,\sqrt{\tau}}\Big)-\mathrm{erf} \Big(-\frac{l+x}{2\,\sqrt{\tau}}\Big)\Big]\end{eqnarray}
is the solution of the linear diffusion equation with the initial condition (\ref{initial1}).
In the limit of small $l$ we obtain that
\begin{eqnarray}\label{mean_sq_t}\Delta(\tau)\,
\approx\frac{\alpha^2\,\theta_{max}^2}{2\,\pi\,\tau}\,l^2\;.\end{eqnarray}

\begin{figure}
\includegraphics[width=0.4\textwidth]{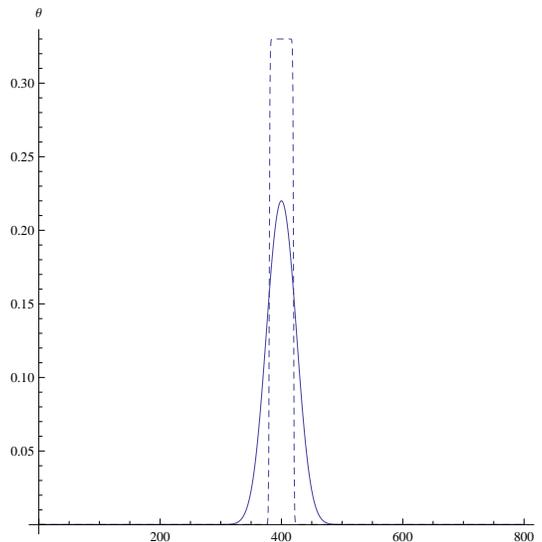}
\caption{Concentration profiles for initial pulse-like distribution (dashed line) and for $\tau=200$  (solid line). The nonlinearity parameter $\alpha=0.2$.}\label{fig:profile_variance}
\end{figure}

\begin{figure}
\includegraphics[width=0.4\textwidth]{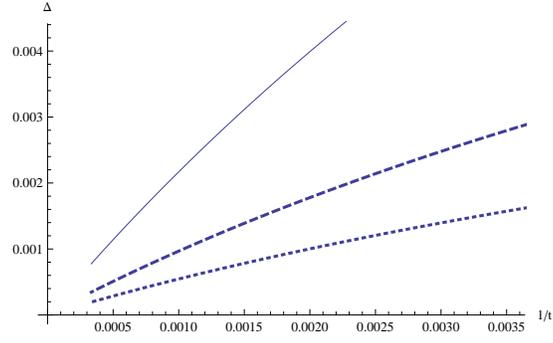}
\caption{Numerically obtained variance rate $\Delta$ given  Eq. (\ref{mean_sq_rate}) as a function of the inverse time $1/t$ for three different values of the nonlinearity parameter $\alpha$: $\alpha=0.15$ (dotted line), $\alpha=0.2$ (dashed line), $~\alpha=0.3$ (solid line).}\label{fig:variance}
\end{figure}

 We checked our analytical considerations by  carrying out numerical simulations of Eqs. (\ref{random_walk_eqn_1D})   with the initial concentration profile given by Eq. (\ref{initial1})(see Fig. \ref{fig:profile_variance})  for different values of the nonlinearity parameter $\alpha$. The results of these simulations are presented in Fig. \ref{fig:variance}. The figure  shows that  the numerically evaluated temporal behavior of the rate function $\Delta$ is in a good agreement with our analytical expression given by Eq. (\ref{mean_sq_t}). Moreover the slopes of the curves  as it is prescribed by the analytics relate as $0.51:0.91:2.0:3.6\approx \alpha_1^2:\alpha_2^2:\alpha_3^2:\alpha_4^2=0.15^2:0.2^2:0.3^2:0.4^2\;.$ Thus, by measuring the temporal behavior of concentration profiles, it is possible to  estimate the strength of interatomic interactions.

\section{Concentration profiles with plateau}
In general, the  diffusion coefficient is a non-monotonic function of atomic  concentration (see e g \cite{naumovets2005}).
There is a number of physical reasons which can cause a non-monotonic coverage dependence of the diffusion coefficient. In thermodynamic terms, the diffusion flux is proportional to the gradient of chemical potential of adsorbed particles $\mu$  which can be written as \cite{adamson},\cite{loburets}
 \begin{eqnarray}\label{mu}\mu=\mu_0-q(\theta)+\frac{1}{\beta}\,\ln\Big(\frac{\theta}{1-\theta}\Big)\end{eqnarray}
The first term in this equation is the standard chemical potential of the adsorbate, $q(\theta)$ is the differential heat of adsorption and the third term stems from the entropy of mixing of adatoms with the vacant adsorption sites on the substrate. (Note that this simplified expression relates only to the first monolayer and does not take into account the possibility of formation of the second and next monolayers).
The diffusion coefficient can be represented as a product
\begin{eqnarray}\label{Dif}D(\theta)=D_j\,\beta\,\frac{\partial \mu}{\partial\,\ln\theta}\;,\end{eqnarray}						
where $D_j$ is a so-called  kinetic factor (or jump diffusion coefficient) \cite{gomer},\cite{naumovets2005},\cite{loburets} and the derivative in the brackets is named the thermodynamic factor. In a simplest case when the cross-correlations between the velocities of diffusing particles are absent, $D_j$ coincides with the tracer diffusion coefficient $D*$ given by Eq. (10).
Inserting Eq. (\ref{mu}) into Eq. (\ref{Dif}), we get
\begin{eqnarray}\label{Diff}D(\theta)=D_j\,\Big(-\beta\,\theta\,\frac{\partial q}{\partial \theta}+\frac{1}{1-\theta}\Big)\;.\end{eqnarray} 					 
It is seen from Eq. (\ref{Diff})  that  any effect which entails a sharp decrease in the heat of adsorption as a function of coverage will result in a maximum of the diffusion coefficient in this coverage range \cite{note}.  For instance, such a situation occurs when all energetically profitable sites at the surface are occupied and adatoms start to fill less favorable sites. Actually, Bowker and King \cite{bowker} found in their Monte Carlo simulations that a well-pronounced  maximum in the $D(\theta)$ dependence observed by Butz and Wagner \cite{butz}  can be explained by existence of two types of lateral interactions: repulsive one between the nearest neighbors and attractive between next-nearest neighbors. A similar effect is typical for volume diffusion of interstitial atoms in disordered binary alloys having a BCC structure with two nonequivalent interstitial positions  \cite{smirnovaa}. In the framework of local equilibrium statistical operator approach \cite{zubarev} it was shown that the physical reason for a non-monotonic concentration dependence coefficient is a combined action of lateral interaction and adatom density fluctuations \cite{tarasenko}.
A sharp drop in the heat of adsorption is also observed in the transition from filling the first, strongly bound (chemisorbed) monolayer to filling the second, weakly bound (e.g., physisorbed) monolayer. In such a case, the spreading of the first monolayer proceeds through diffusion in the mobile uppermost (second or next) monolayer (the so called "unrolling carpet" mechanism) \cite{gomer}. This example shows that a change in the heat of adsorption can be accompanied not only by variation of the diffusion parameters (the activation energy and prefactor $D_0$  in the Arrhenius equation), but also by a change in the atomistic diffusion mechanism itself.

\begin{figure}
\includegraphics[width=0.4\textwidth]{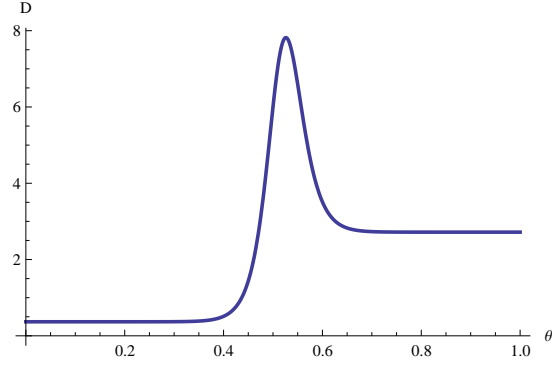}
\caption{Diffusion coefficient for a step-like on-site energy. The parameters used are $~a=1,~\ell=0.05,~\theta_0=0.5$ }\label{fig:diffusion_coeff}
\end{figure}

It is worth noting  also that the non-monotonic concentration dependence  may be phenomenologically connected with a step-like dependence of the heat of adsorption  on the coverage ( see a review paper \cite{naumovets89}) . Fig. \ref{fig:diffusion_coeff} shows the diffusion coefficient calculated from Eq. (\ref{nonl_diff_coeff}) by assuming that the on-site  adatom energy ${\cal E}(\theta)$ (which in most cases  is proportional to the heat adsorption $q(\theta)$) has the form
\begin{eqnarray}\label{on_site_step}
{\cal E}(\theta)=\alpha\,\Big(1+\tanh\kappa\,(\theta-\theta_{thr})\Big)\end{eqnarray}
where the parameter  $\theta_{thr}$ gives the threshold value of the coverage and  the parameter $\kappa$ characterizes the sharpness of the transition to the new state \cite{naum}.

The aim of this section is to consider  the diffusion process with a step-like concentration dependent on-site energy given by Eq. (\ref{on_site_step}) and clarify what kind of new information one can derive  by comparing  theoretically obtained concentration  profiles with experimentally obtained ones.

\begin{figure}
\includegraphics[width=0.4\textwidth]{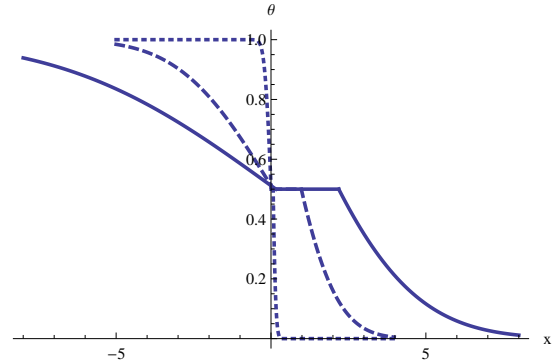}
\caption{Analytically obtained concentration  profile for the diffusion coefficient in the  $\delta$-function limit (\ref{on_site_step})   with $\alpha=1,~\theta_0=0.5,~\Theta=1$ for three different time moments : $~\tau=0.005$ (dotted line), $~\tau=1$ (dashed line), $~\tau=5$ (solid line).}\label{fig:front_anal}
\end{figure}

\begin{figure}
\includegraphics[width=0.4\textwidth]{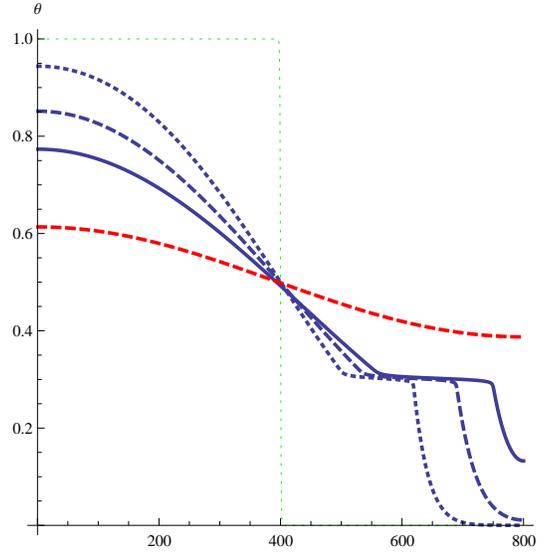}
\caption{(color on-line) Numerically obtained concentration  profiles in the  case of the step function energy dependence given by Eq. (\ref{on_site_step}) with $\theta_0=0.5,~~\alpha=1,~~\ell=0.005$  for different time moments: $t=0$  (dotted gray line), $t=6000$ (dotted line), $t=12000$ (dashed line), $t=20000$ (solid line),$~t=40000$ (dashed, red, thick line). }\label{fig:front_numer_ext}
\end{figure}

 It is very hard and may be hopeless to solve the equation (\ref{nonl_diff_eq})  with the diffusion coefficient given by Eqs. (\ref{nonl_diff_coeff}) and (\ref{on_site_step}). However, the problem can be solved and some insight into the kinetics can be achieved in the limiting case of very sharp energy concentration dependence: $~\kappa\rightarrow \infty\;.$ In this case the diffusion coefficient  (\ref{nonl_diff_coeff}) and (\ref{on_site_step}) takes the form
 \begin{eqnarray}\label{nonl_diff_coeff_step}
 D(\theta)=D^{*}\,\Big(1+a\,\delta(\theta-\theta_{thr})+b\,H(\theta-\theta_{thr})\Big)\end{eqnarray}
 where \begin{eqnarray}\label{cs}a=\alpha\,e^{\alpha}\theta_{thr}\,(1-\theta_{thr}),
 ~~b=e^{2\,\alpha}-1.\end{eqnarray}
The nonlinear diffusion equation (\ref{nonl_diff_eq}) with the diffusion coefficient (\ref{nonl_diff_coeff_step}) and the initial condition (\ref{initial}) has a self-similar solution
$\theta(x,\tau)\equiv\Theta(z),~(z=x/2\sqrt{\tau})$  which can be presented in the form (see Appendix for a detailed derivation)
 \[\theta(x,\tau)=\left\{ \begin{array}{rl}\theta_{thr}\,\frac{\mathrm{erfc}\Big(\frac{x}{2\,\sqrt{\tau}}\Big)}
 {\mathrm{erfc}\big(z_1\big)}\,, &\mbox{when}~~x\geq 2\,z_1\,\sqrt{\tau}\\
\theta_{thr}\,\;,&\mbox{when}~~-\,2\,z_2\,\sqrt{\tau}\,e^{\alpha} \leq x \leq 2\,z_1\,\sqrt{\tau}\\
\theta_{max}-\Big(\theta_{max}-\theta_{thr}\Big)
\,\frac{\mathrm{erfc}\Big(-\frac{x}{2\,\sqrt{\tau}}\,e^{-\alpha}\Big)}{\mathrm{erfc}\big(z_2\big)}\;, &\mbox{when}~~x\leq -2\,z_2\,e^{\alpha}\,\sqrt{\tau}
\end{array}\right.\]
Here the parameters $z_1$ and $z_2$ are determined by the equations
\begin{eqnarray}\label{z12s}z_2=\frac{\sqrt{\pi}}{2}\,\alpha\,\,
(1-\theta_{thr})
 \,e^{z_1^2}\,\mathrm{erfc}(z_1)-\,z_1\,e^{-\alpha}\;,\nonumber\\
e^{\alpha}\,\Big(\theta_{max}-\theta_{thr}\Big)\,e^{z_2^2}\,\mathrm{erfc}(z_2)=
\theta_{thr} \,e^{z_1^2}\,\mathrm{erfc}(z_1)
\end{eqnarray}
which are obtained from Eqs. (\ref{z12}) taking into account the definition (\ref{cs}).
 Thus, the concentration profile in the case of non-monotonic diffusion coefficient  is characterized by existence of a plateau where the concentration of adatoms does not depend on the spatial variable $x$. The length of the plateau  $\ell_p=2\sqrt{\tau} (z_1+z_2\,e^{\alpha})$ increases with time. Such a behavior is shown in Fig. \ref{fig:front_anal}. The rate with which the length $\ell+p$ of the plateau increases is determined by the nonlinear parameter $\alpha$ and the threshold coverage $\theta_{thr}$.
We carried out also numerical simulations of Eqs. (\ref{random_walk_eqn_1D}) with the the step-like on-site energy ${\cal E}_n=\alpha\,\Big[1+\tanh\Big(\kappa\,(\theta_n-\theta_{thr})\Big)\Big]$   and as it is seen from Fig. \ref{fig:front_numer_ext}  our simple model (\ref{nonl_diff_coeff_step}) is in reasonable agreement  with numerics. Note that the plateau in the concentration dependence develops only on intermediate stage of the evolution.  For large enough times the height of the concentration profile becomes small,  effects of interatomic interactions are negligible and the profile evolutes in accordance with the linear diffusion equation (see Fig. \ref{fig:front_numer_ext}).

It is worth noting that  the diffusion of Dy adatoms absorbed by Mo (1 1 2) for the initial coverage $\theta(x,0)\approx 0.7 H(-x)$ shows a very well pronounced plateau in the concentration profile dependence both on the spatial coordinate for different time moments   and on  the Boltzmann variable ( see Figs. 7 and 8 in \cite{nikitin}).  It means that as it is prescribed by our analytical considerations  the length of the plateau increases as $t^{1/2}$. This suggests that our simple analytical model may be a useful tool in analyzing  an experimentally observed  concentration behavior.
\section{Conclusions and discussion}
In this paper, we have investigated the role of interactions between adatoms in surface diffusion.
The problem was considered analytically in the mean-field approach.
By analyzing  discrete nonlinear random walk equations and corresponding nonlinear diffusion equations with an initial condition in a form of step-like concentration profile, we have found that the interactions between adatoms  influence significantly the concentration profile development on early and intermediate stage of the process. In the case of low coverage the  interaction between adatoms makes the concentration profile asymmetric: it is shifted to the side of high concentration in the case of repulsive interactions and to the side of low concentration for attractive interactions. By calculating the magnitude of the shift one can estimate the intensity of lateral interactions between adatoms.  At the late stage of kinetics the role of interatomic interactions becomes negligible.
By studying the  nonlinear random walk process which is characterized by a sharp maximum in the concentration dependence of the diffusivity, we have found that  a  well-pronounced plateau develops in the concentration profile. The length of the plateau increases in time as $t^{1/2}$ . The height of the plateau  $\theta_{thr}$  corresponds to the maximum of the diffusion coefficient which in the frame of our approach corresponds to a sharp decrease in the heat of adsorption as a function of coverage. The rate with which the length of the plateau increases with time   is determined by an amount at which the adsorption heat drops at the threshold coverage $\theta_{thr}$ . All above mentioned results can be verified experimentally.

\section*{Acknowledgements} The authors acknowledge   support from a Goal-oriented
program of the National Academy of Sciences of Ukraine.

\appendix
\section{} \label{appendA}

The nonlinear diffusion equation (\ref{nonl_diff_eq}) with the diffusion coefficient (\ref{nonl_diff_coeff_step}) and the initial condition (\ref{initial}) has a self-similar solution
$\theta(x,\tau)\equiv\Theta(z),~(z=x/2\sqrt{\tau})$  which satisfies the equation
\begin{eqnarray}\label{nonl_diff_eq_b}-2\,z\,\frac{d\Theta}{d z}=\frac{d}{d z}\Big(D(\Theta)\,\frac{d\Theta}{d z}\Big)\;.
\end{eqnarray}
The boundary conditions for Eq. (\ref{nonl_diff_eq_b}) are
\begin{eqnarray}\label{bound_cond}\Theta(z)\rightarrow\Theta,~~~\text{for}~~z\rightarrow-\infty\;,\nonumber\\
\Theta(z)\rightarrow 0,~~~\text{for}~~z\rightarrow\infty\;.\end{eqnarray}

From Eqs. (\ref{nonl_diff_eq_b}) and (\ref{nonl_diff_coeff_step}) we see that  the function
\begin{eqnarray}\label{y}y(\Theta)=\int\limits_0^\Theta\,d\Theta'\,z(\Theta')\;,\end{eqnarray}
where $z(\Theta)$ is an inverse function with respect to $\Theta(z)$,  satisfies the equation
\begin{eqnarray}\label{eq_y}-2\,\frac{d^2 y}{d \,\Theta^2}=D(\Theta)\frac{1}{y} \end{eqnarray}
or equivalently, two equations
\begin{eqnarray}\label{eqs_y}-2\,\frac{d^2 y}{d\,\Theta^2}=\frac{1}{y},~~~~\text{for}~~\Theta < \theta_{thr}\;,\nonumber\\
-2\,\frac{d^2 y}{d\,\Theta^2}=\frac{1+b}{y},~~~~\text{for}~~\Theta > \theta_{thr}\;
 \end{eqnarray}
 augmented by the jump   condition
\begin{eqnarray}\label{jump} \frac{d y}{d\Theta}\Big|_{\Theta=\theta_{thr}+0}-\frac{d y}{d \Theta}\Big|_{\Theta=\theta_{thr}-0}=
-\frac{a}{2\,y(\theta_c)}\;,\end{eqnarray}
and the continuity condition
\begin{eqnarray}\label{cont} y(\theta_{thr}+0)=y(\theta_{thr}-0)=
y(\theta_c)\;.\end{eqnarray}
By integrating Eqs. (\ref{eqs_y}) once, we get
\begin{eqnarray}\label{eqs_y_i} \frac{d y}{d \Theta}=\sqrt{2 z_1+\ln\frac{y(\theta_{thr})}{y}},~~~~\text{for}\,\,\,\Theta < \theta_{thr}\;,\nonumber\\
-\frac{1}{\sqrt{1+b}}\frac{d y}{d \Theta}=\sqrt{2\,z_2+\ln \frac{y(\theta_{thr})}{y}},~~~~\text{for}\,\,\,\Theta > \theta_{thr}\;,\end{eqnarray}  where the constants $y(\theta_{thr})$, $z_1$ and $z_2$ satisfy the equations
\begin{eqnarray}\label{z12}\theta_{thr}=\frac{\sqrt{\pi}}{2}\,\frac{a}{z_1+\sqrt{1+b}\,z_2} \,e^{z_1^2}\,\mathrm{erfc}(z_1)\;,\nonumber\\
\sqrt{1+b}\,\Big(\theta_{max}-\theta_{thr}\Big)=\frac{\sqrt{\pi}}{2}\,\frac{a}{z_1+\sqrt{1+b}\,z_2} \,e^{z_2^2}\,\mathrm{erfc}(z_2),\nonumber\\
 y(\theta_{thr})=\frac{a}{z_1+\sqrt{1+b}\,z_2}
\end{eqnarray}
which were obtained from the jump condition (\ref{jump}) and the continuity condition (\ref{cont}). Taking into account the definition (\ref{y}), we  obtain eventually from Eqs. (\ref{eqs_y}) that the concentration profile is determined by the following expressions
\begin{eqnarray}\label{theta_sol} \theta(z)=\theta_{thr}\,\,\frac{\mathrm{erfc}(z)}{\mathrm{erfc}(z_1)} ~~~~~\text{when}~~z\geq z_1\;,\nonumber\\
\theta(z)=\theta_{thr}~~~~~\text{when}~~-\sqrt{1+b}\,\,z_2 \leq z \leq z_1\;,\nonumber\\
\theta(z)=\theta_{max}-\Big(\theta_{max}-\theta_{thr}\Big)\,\,\frac{\mathrm{erfc}
\Big(-\frac{z}{\sqrt{1+b}}\Big)}{\mathrm{erfc}(z_2)} ~~~~~\text{when}~~z\leq -z_2\,\sqrt{1+b}\;.
\end{eqnarray}


\begin{thebibliography}{14}
\bibitem{note}It is worth noting that Eq. (\ref{Dif}) is valid for $\theta\,<\,1$ and the divergence $D(\theta)$ for $\theta\rightarrow 1$ is a model effect. It does not appear when the possibility of filling the second and next monolayers is taken into consideration.
\end{thebibliography}
\end{document}